\documentclass[12 pt]{article}
\usepackage{graphicx}
\usepackage{amssymb}
\usepackage{amsmath}
\usepackage{amsthm}

\begin{document}

\title{A Local Approach for Information Transfer}

{\tiny
\author{P. Garc\'{\i}a$^a$\footnote{Permanent address: Laboratorio de Sistemas Complejos, Departamento de F\'{\i}sica Aplicada, Facultad de Ingenier\'{\i}a, Universidad Central de Venezuela,
Caracas, Venezuela.} and R. Mujica$^b$\\
$^a$Facultad de Ingenier\'{\i}a en Ciencias Aplicadas,\\
 Universidad T\'ecnica del Norte,
Ibarra, Ecuador.\\ 
$^b$ Centro de F\'{\i}sica Te\'orica y Computacional,\\ Laboratorio de Fen\'omenos No Lineales,\\
Escuela de F\'{\i}sica, Facultad de Ciencias,\\ Universidad Central de Venezuela,\\
Caracas, Venezuela.}
}

\date{}

\maketitle
\thispagestyle{empty}

\abstract{In this work, a strategy to estimate the information transfer between the elements of a complex system, from the time series associated to the evolution of this elements, is presented. 
By using the nearest neighbors of each state, the  local approaches of the deterministic dynamical rule generating the data and the probability density functions, both marginals and conditionals,  necessaries to calculate some measures of information transfer, are estimated. \\
The performance of the method using numerically simulated data and real signals is exposed.
}

\section{Introduction}
\setcounter{equation}{0}
The estimation of directionality of coupling between electrocardiographic and respiratory signals\cite{Schultz} may help to decide on which system exerts control allowing a more efficient medication; determination of the causal relations between shares of stock, to select which one to monitor, in order to make decisions to buy or sell; the detection of the elements that control the evolution of the others in a complex network, allows to estimate its robustness and simplifies the control problem by reducing the number of sites on which to take control actions\cite{Liu}; these are some of the situations where it is useful to estimate how a subsystem sends and receives information from another in a complex system.

From a more general perspective, the estimation of the extent to which a component contributes to the production of information system and the rate at which it is shared with the rest of the components, can provide important information about its structure. For this reason, detecting the transfer of information and its directionality is a subject of great interest due to the variety of practical applications, in areas ranging from physics to marketing.

Since Wiener\cite{Wiener} proposed that an improvement of the prediction of the future of a time series $x$, by the incorporation of information from the past of a second time series $y$, can be seen as an indicator of a causal interaction from $y$ to $x$, 
an operational way to implement these ideas was proposed by Granger \cite{Granger} and a further formalization of the strategy by 
Schreiber \cite{Schreiber}; the problem of the estimation of a cause-effect relation between elements of a complex system, have been 
addressed under many different points of views. 
From deterministic \cite{Bezruchko,Ma,Ishiguro,Cisneros,Rosenblum} to probabilistic \cite{Schreiber,Palus,Vejmelka} 
modeling schemes have been proposed and the problem has been named as detection of causality \cite{Ma,Ishiguro} or  directionality of the coupling 
estimation \cite{Bezruchko,Rosenblum,Palus,Vejmelka} or detection of direct links \cite{Rubido}, among others. 

In this work we propose to estimate the transmission of information between elements of complex systems, using the formal probabilistic scheme proposed by ​​Schreiber in \cite{Schreiber}, but approximating the probability density functions (PDF) involved, by using a $k$-nearest neighbors approach.  In this case, the neighborhood, to any point, is used in a  deterministic nonlinear local model to estimate the conditional PDF's and  in a density estimator to approximate the marginals PDF's associated with the data.  This strategy, although it is also based on the determination of the $k$ nearest neighbors, is conceptually simpler than the strategy presented in \cite{Zhu} and has the advantage that the estimate can be made using only the nearest neighbor, which in principle makes it computationally less expensive. 

The use of deterministic models for the further determination  of the  information transfer, is expected to offer an interesting relationship between the strategies based on probabilistic and deterministic models, as well as between the parametric and non-parametric schemes.

In order to show the usefulness of the scheme, the work is organized as follows: section 1 is devoted to present the idea behind of the information transfer and section 2 gives a methodology to estimate the conditional PDF's involved in information transfer, using local approaches. In section 3 we combine these ideas, into the approach of Schreiber \cite{Schreiber}, to estimate the information transfer in the case of numerically simulated and real signals. Finally in section 4, we give some concluding remarks. 

\section{Information transfer}
\label{section1}
There are many tools  to estimate the information transfer in time series, see for example \cite{Vicente} and references therein. Nevertheless an attempt to classify them in few 
categories require a titanic effort, among other reasons,  because of the unclear relations between them. However, most popular time series tools appear to fall into one of two broad classes, model-based strategies and parameter-free strategies or model-free strategies.

The representative of the first class is an approach called Granger Causality \cite{Granger} and the other class is represented 
by a scheme based on information theory, entropy transfer, proposed by Schreiber in \cite{Schreiber}.

In Granger's method, causality could be tested by measuring the ability to predict, using a linear model, the future values of a time series using prior values of another time series. 

Using this scheme in the case of nonlinear subsystems can leads to erroneous conclusions. This problem is addressed in  \cite{Ancona} where a nonlinear model is performed using radial basis functions. Despite this, the scheme still has the weakness of being a static measure.

In the case of parameter-free methods,  the usual strategy to quantify the superimposing of information contained in 
interacting 
subsystems is the mutual information. Unfortunately this is a symmetrical and static  measure, so that does not give a sense of 
direction or temporal evolution of information in the system. 
A strategy which overcomes the two previously mentioned weaknesses is proposed in \cite{Schreiber} to estimate the transmission of 
information based on observations of the time series associated with the system elements and information theory \cite{Shannon}.

In \cite{Schreiber}, Schreiber proposes a measure of transfer of entropy which represents a rigorous derivation of Wiener causal measure 
and shares some of the desired properties, to represent transfer of information,  with Mutual Information but takes into account 
the dynamics of the system.  

This measure, with minimal assumptions about the dynamics of the system and the nature of the coupling, is able to quantify the 
exchange of information between two systems $x$ and $y$ as follows: 

\begin{eqnarray}
I_{x \to y} &=&  \sum_{y_{n+1},x_{n},y_{n}} p(y_{n+1},y_{n},x_{n}) \log_2 \left( \frac{p(y_{n+1}|y_{n},x_{n})}{p(y_{n+1}|y_{n})} \right)\nonumber \\
I_{y \to x} &=&  \sum_{x_{n+1},x_{n},y_{n}} p(x_{n+1},x_{n},y_{n}) \log_2 \left( \frac{p(x_{n+1}|x_{n},y_{n})}{p(x_{n+1}|x_{n})} \right)
\label{ti}
\end{eqnarray}

\noindent
and it may be thought \cite{Palus-2001} as a {\it conditional mutual information}.

Where $p(\cdot,\cdot)$ and $p(\cdot|\cdot)$ are the joint and conditional probabilities of occurrence of the state or system. 
Determining the flow of information given by (\ref{ti}) requires the calculation of probability densities associated with 
transitions between states of each subsystem, and in this  case of a brute-force estimation, requires a coarse graining of the
time series associated to both subsystems. This represents a large computational cost that increases considerably when these probabilities should be calculated for all pairs of elements of a large extended system.

\section{Probability densities and local modeling}
According to Wiener's idea, if the improvement in prediction can be associated with a reduction in uncertainty, it is natural that a measure of causality can be represented in terms of information theory concepts. However, the determination of causal relationships or information transfer, may also be represented in terms of deterministic models.

To show the feasibility of the idea, we start by representing the joint probabilities, in the entropy transfer definition (\ref{ti}), in terms of conditional and marginal probability densities using: $ p(x,y,z) = p(x|y,z) p(y|z) p(z) $,  for later,  we     propose a strategy to estimate both, the conditional densities probabilities and marginal densities probabilities, using local nonlinear modeling of the dynamical system \cite{Farmer,Garcia-1996,Garcia-2007} and nonparametric method to estimate  marginals densities functions \cite{Loftgaarden} respectively. 

Inspired in \cite{Husmeier}, we propose to estimate the conditional probabilities densities 
$p(x_{n+1}|x_n)$,  ~$p(x_{n+1}|x_n,y_n)$,  ~$p(x_{n+1}|y_n)$  and  ~$p(y_{n+1}|y_n)$, ~ $p(y_{n+1}|y_n,x_n)$,  $p(y_{n+1}|x_n)$ associated to the appearance of  states $x_{n+1}$ and $y_{n+1}$, given some of the predecessors states of both subsystems, using the quality of predictions made from a deterministic model.
Specifically, we base the approach in the estimation of the nearest neighbors of $x_n$ and $y_n$. The marginal probability density functions, $p(x_n)$ and $p(y_n)$,  can be approached using the same strategy.

In order to do this, we start by constructing a Cumulative Distribution Function to derive from it the necessary probability densities. This function is, by definition,  a real-valued and strictly increasing  function of  a random variable $X$, usually represented as:

\begin{equation}
F_X(x) = P(X \leq x),
\label{distribution}
\end{equation}

\noindent
where the right-hand side is the probability that, the random variable $X$ takes on a value less than or equal to $x$. The probability that $x$ lies in the semi-closed interval $(a, b]$, where $a < b$, is therefore

\begin{equation}
P(a < X \le b)= F_X(b)-F_X(a). 
\end{equation}
 
The Cumulative Density Function of a continuous random variable $X$ can be expressed as the integral of its probability density function $p_X(x)$ as follows:

\begin{equation}
F_X(x) = \int_{-\infty}^x p_X(t)\,dt.
\end{equation}

Similarly, for the case of Cumulative Conditional Distribution Function:

\begin{equation}
F_{XY}(y|x) = \int_{-\infty}^y p_{XY}(y|x)\,dy.
\label{CCDF}
\end{equation}

Thus, the conditional probability density (CPD) function can be obtained as:

\begin{equation}
p_{XY}(y|x) = \frac{\partial F(y|x)}{\partial y}
\label{distribution}
\end{equation}

\noindent
If we chose the Cumulative Conditional Distribution Function in (\ref{distribution}) as a sigmoid function, $S(u)=1/1 + e^{-r u}$, with $u=f(x)-y$, as in \cite{Husmeier}:

\begin{equation}
F(y|x) = S(f(x) - y)
\label{determ}
\end{equation}

\noindent
where $f$ is a model for the system, constructed from the data;  then we can distinguish between two extremes for distribution: one strictly deterministic, i.e. without errors, where $r \to \infty$ and $F$ is given by:

\begin{equation}
F(y|x) = \theta (f(x) - y),
\label{determ}
\end{equation}

\noindent
with $\theta$ the Heaviside function, or a probabilistic approach, where the parameter $r < \infty$. 
Here $r$ is, in some sense, a measure of the randomness of the process $X$. 

It is clear, from equations (\ref{distribution}) and (\ref{determ}), that if we have an approximation of the dynamic rule that governs the evolution of the system, then it is possible to construct an adequate cumulative distribution function, that allows to estimate the probability densities, necessary to calculate the transfer of information. In this case we propose to chose the parameter $r$, in the sigmoid function, proportional to $1/\sigma$, where  $\sigma$ is the standard deviation of the modeling errors, since the slope of the transition of the sigmoid function determines the width of the distribution associated with it. 

At this point, we can summarize the presentation into two main ideas:  the Schreiber's information transfer, and some other estimators, can be written in terms of CPD's and  the CPD's can be estimated using deterministic models of the rule generating the data. The remainder of the section, is intended to implement these models and show how they can be incorporated into an estimator of information transfer,  with a computational cost similar to the determination of the $k$-nearest neighbors, with $k = 1$. 

Let us suppose that we have a series of values of the states of a dynamical system  $\{x_i \}_{i=1}^N$, with $x_i \in R^d$, obtained either by measuring the $d$ components of system's  state at regular time intervals or by reconstructing the state space by using the Takens theorem\cite{Takens}, from partial information about the states. 

In our approach the dynamical rule is estimated using a local approximation for $f$. This approach can be presented algorithmically as:

\begin{itemize}
\item[i.] Given the $i$-th data value $x_i$, with $i=1,2,\cdots,N$.
\item[ii.] Determine the set of the $m$ nearest states  ($\{x_{v^i_j} \}_{j=1}^m$), to the state $i$ in the Euclidean metric, sorted in ascendant order with the distance from $x_i$. 
Here  $v^i_j$, the label of the $j$-th neighbor of the $i$-th data value, is an integer in the interval $[1, N-1]$. 
\item[iii.] The approximation is obtained by Taylor expanding $f$  to first order, around of the nearest neighbor   $x_{v^i_1}$,   to  obtain:

\begin{equation}
x_{i+1} \approx f(x_{v^i_1}) + Df(x_{v^i_1}) (x_i - x_{v^i_1})
\end{equation}

where, $ Df(x_{v^i_1})$ is the Jacobian matrix of $f$.
\item[iv.] Finally, in zero order, the evolution of the system is approximated by the evolution of the nearest neighbor, i.e.,  $x_{i+1} \approx x_{v^i_1+1} $. For a first order approximation it is necessary to calculate $ Df(x_{v^i_1})$.
\end{itemize}

The marginal probabilities densities functions $p(x)$ y $p(y)$, can be estimated using the $k$-th nearest neighbor density estimator\cite{Loftgaarden}:
  
\begin{equation}
p(x_i) = \frac{k}{N ||x-x_{v^i_k}||}.
\label{MPDF}
\end{equation}

\begin{figure}[h]
\centerline{
\includegraphics[width=0.5\textwidth]{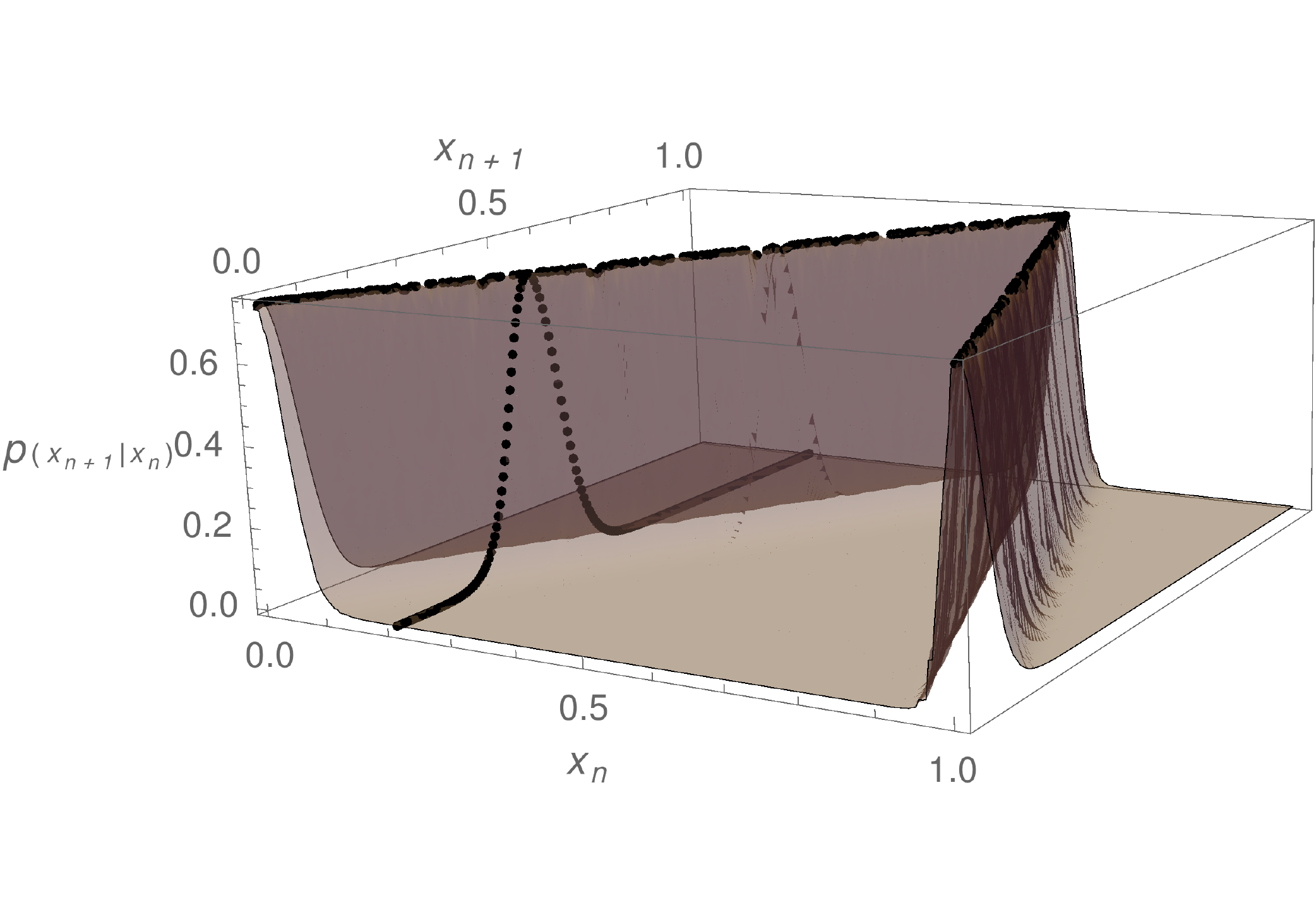}
\includegraphics[width=0.45\textwidth]{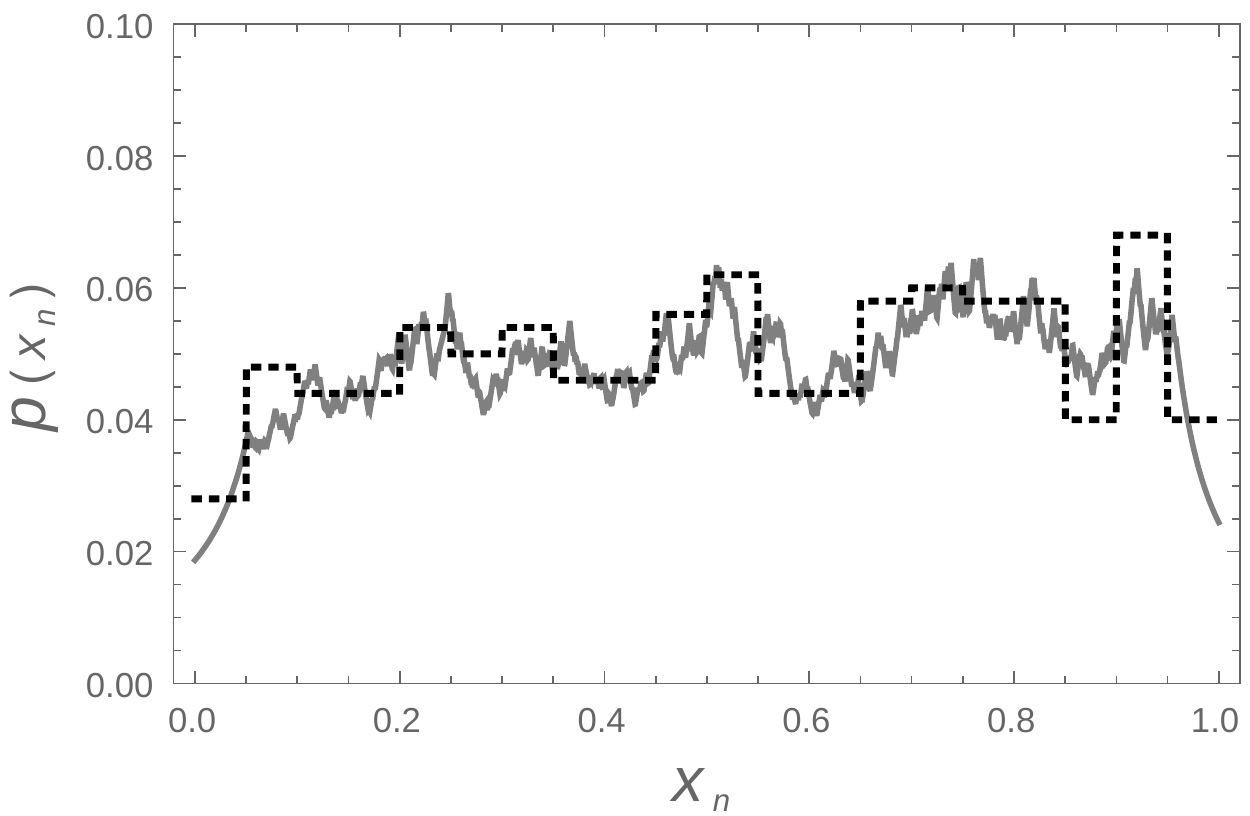}
}

\caption{Left: CPD function. Doted line shows, as an example, the probability of the occurrence of all values given the occurrence of the value 0.2. Right: Marginal probability density function. 
The solid line represents the local estimation and the doted one represents a histogram of the data.  Both PDFs are estimated using 500 data points of the skew tent map with parameter $a=0.65$ and $k=1$.}
\label{f1}
\end{figure}

To illustrate graphically the methodology to estimate the probability densities, Figure \ref{f1} shows conditional and marginal probability density functions,  for $N=500$ data points of the skew tent map with $a=0.65$, 

\begin{equation}
f_a(x_i) = \left\{
\begin{array}{cc}
 \frac{x_i}{a} & 0 \leq x \leq a \\
 \frac{1-x_i}{1-a} & a< x \leq 1 \\
\end{array} \right.,
\label{tent}
\end{equation}

\noindent
Thus, for zero order approach of the dynamical rule $f$ in (\ref{tent}) and using the equations (\ref{distribution}) and (\ref{determ}), the  CPD function is given by:

\begin{equation}
p(x_{i+1}|x_{i})=\frac{r e^{-r(x_{v^i_1 + 1}-x_{1+1})}}{\left[1+e^{-r(x_{v^i_1+1}-x_{i+1})}\right]^2 }
\end{equation}

Figure 1, illustrates the above mentioned ideas and gives a graphical insight of how the proposed strategy associates the modeling errors with CPD's.

\section{Results}
In order to show the performance of our strategy, we estimate the information transfer in the case of signals from a numerical simulation and real data. 

\subsection{Numerical simulations}
The data associated with the numerical simulations correspond to two coupled skew tent maps in chaotic regime, connected by a nonlinear and bidirectional coupling function  as in \cite{Hasler} and the Chuas's system\cite{Chua} the first case provides a controlled experiment where the directionality of the transfer of information is known, which allows a performance testing strategy; the second example will allow to compare the performance of the strategy using numerically  simulated data with the results for the same system but in the case of real data.

\subsubsection{Coupled chaotic maps}
\begin{figure}[h]
\centerline{
\includegraphics[width=0.45\textwidth]{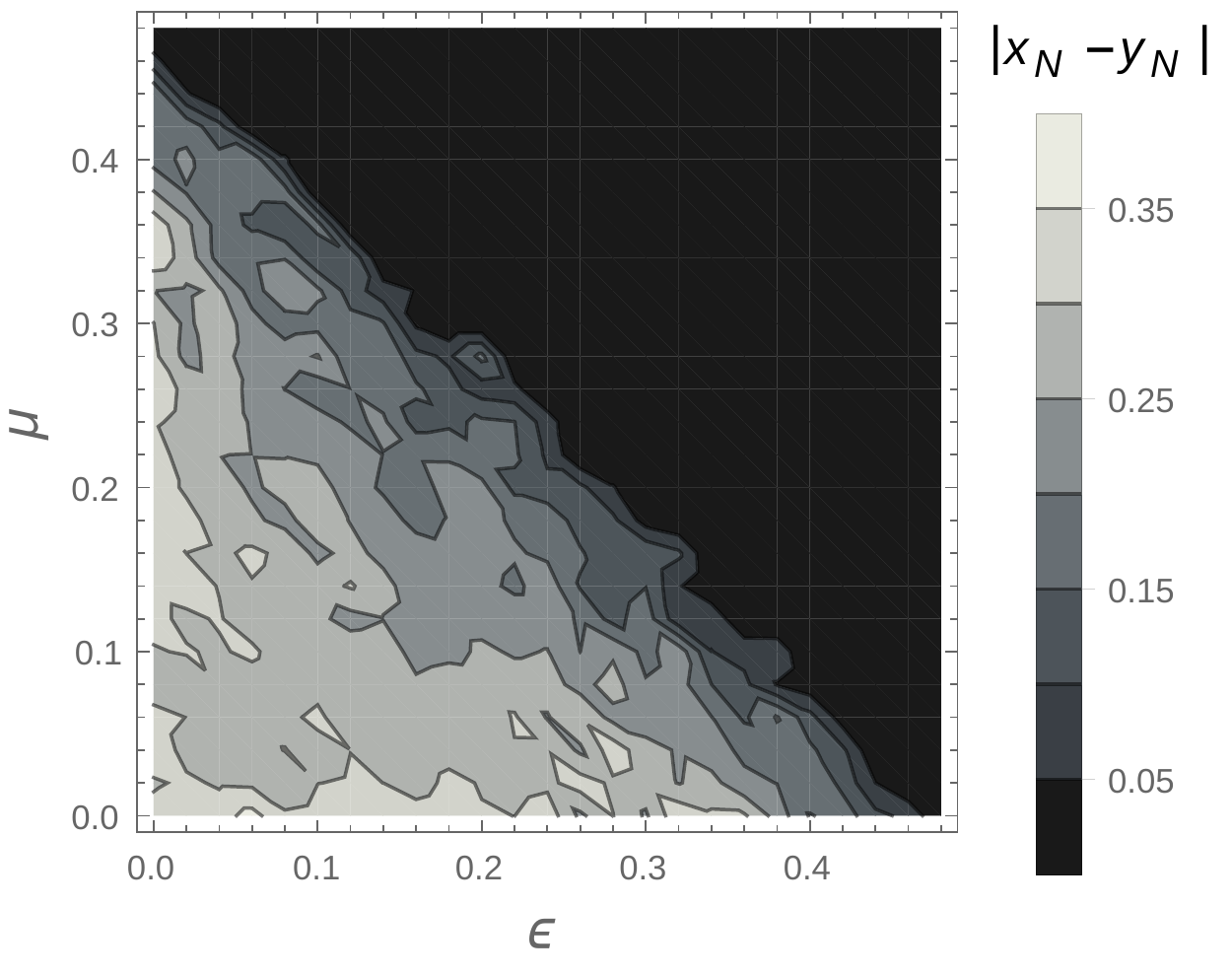}
\includegraphics[width=0.55\textwidth]{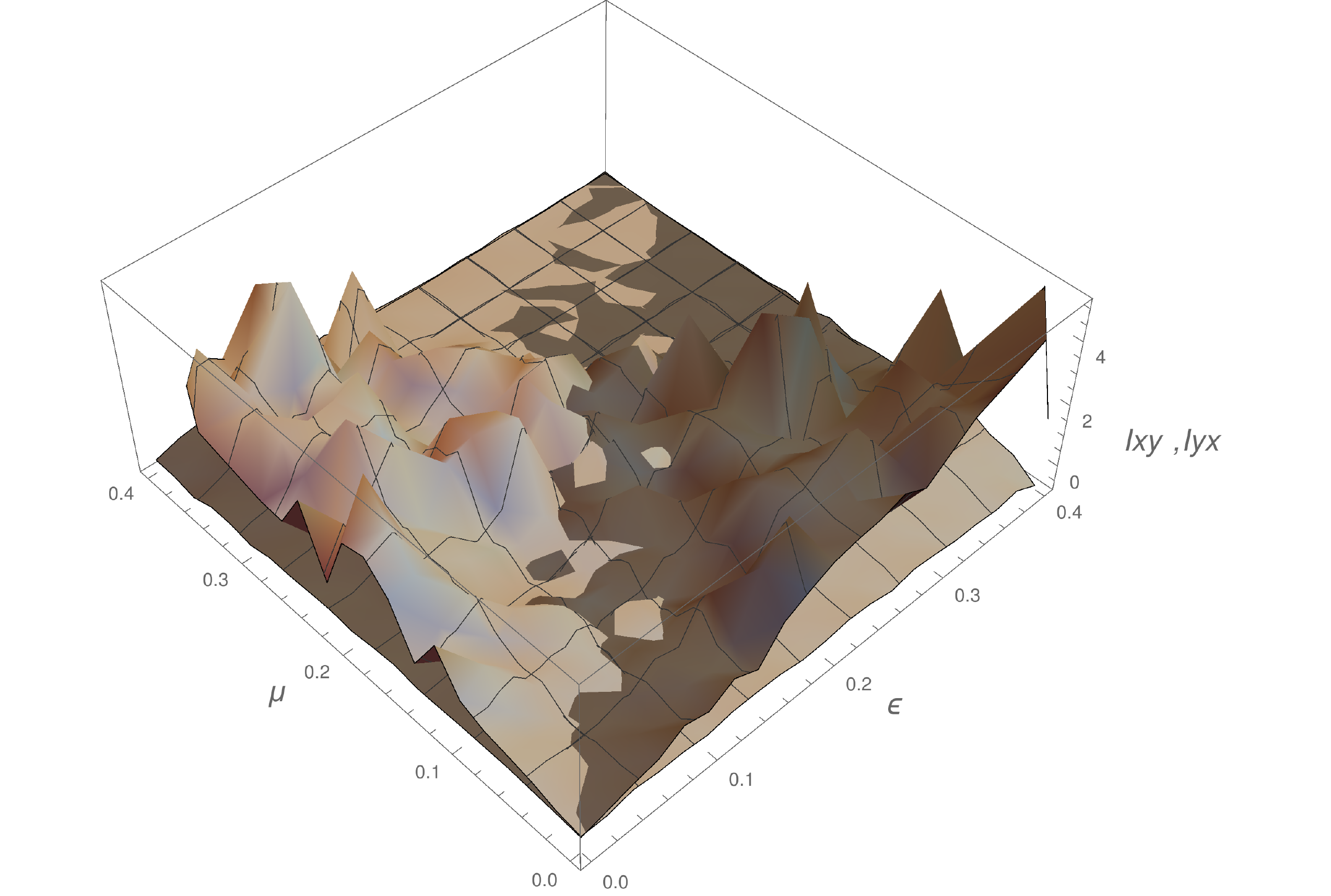}
}
\caption{Left: Synchronization error $|x_N-y_N|$ as a function of the coupling parameters. Right: Information Transfer between the elements of the systems (\ref{coupled maps}), estimated using $500$ data points.}
\label{f2}
\end{figure}

Here, two chaotic tent maps (\ref{tent}) with parameter $a=0.5$, are coupled according to:

\begin{eqnarray}
 x_{n+1} &=& f(x_n + \epsilon (y_n - x_n))  \nonumber \\
 y_{n+1} &=& f(y_n + \mu (x_n - y_n)) 
 \label{coupled maps}
\end{eqnarray}

\noindent
where the parameters $\epsilon$ and $\mu$ define the intensity of the coupling. The transmission of information, in this case, can be associated to the synchronization of the trajectories of the systems \cite{Bollt}. Being understood by synchronization, the coincidence between the states of the subsystems once sufficient time has elapsed.
Figure \ref{f2} shows synchronization error and information transfer between the coupled maps (\ref{coupled maps}) as a function of the coupling parameters. It is worth to note as  the flat zone, in the information transfer surface, coincide with the synchronization zone,  as suggested by Bollt in \cite{Bollt}.

\subsubsection{Chua's system}
The Chua's system is a well-known chaotic system, representing a $RLC$ circuit with one or more nonlinear elements. The easy construction of the circuit has made it an ubiquitous real-world example of a chaotic system.

\begin{eqnarray}
\frac{dV_{C_1}}{dt}   &=& \alpha [V_{C_2}-V_{C_1}-g(V_{C_1})] \nonumber \\
R C_2 \frac{dV_{C_2}}{dt}  &=& V_{C_1} - V_{C_2} + R i_L  \nonumber \\
\frac{di_L}{dt} &=&-\beta V_{C_2}
\label{chua-system}
\end{eqnarray}

\noindent
where $g(V_{C_1})$, is the usual piece-wise linear function representing the Chua's  diode and all parameters are chosen as in \cite{Chen-a}.

\begin{figure}[h]
\centerline{
\includegraphics[width=0.45\textwidth]{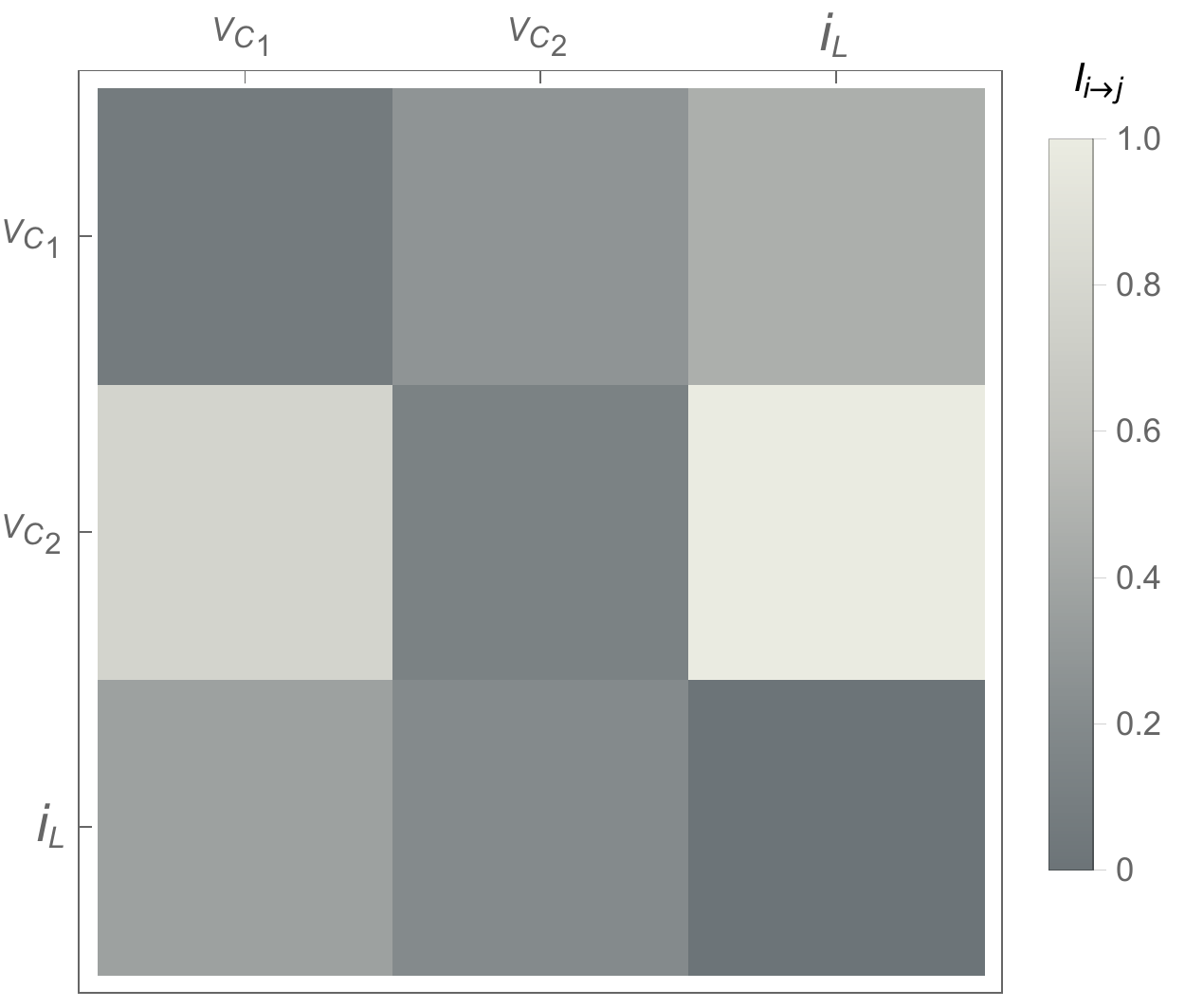}
}
\caption{Information transfer between all pairs formed with signals $V_{C_1}$, $V_{C_2}$ and $i_L$. The data is obtained by numerical integration of the system (\ref{chua-system}) using a Runge-Kutta of $4$-th order. Here we chose the parameter $r$ as two times the standard deviation of the modeling errors at dimension three.}
\label{f3}
\end{figure}

Let us rename the elements of the circuits indexing them as $C_1 \to 1$, $C_2 \to 2$ y  $L \to 3$ and define the net flow of information associated to each component of the circuit, as the difference among the sum of the information transfer between it and the rest of the elements and the sum of the information transfer from the rest of the elements to it, as:

\begin{equation}
 T_i = \sum_{j=1}^3 I_{i \to j} - I_{j \to i}. 
 \label{inf-net}
\end{equation}

\noindent
The evaluation of these quantities, see Figure \ref{f3}, produces: 
$ T_1=-0.5684, T_2=1.6476, T_3=-1.0792$. This is consistent with some results about the control of Chua's circuit \cite{Chen-a,Chen-b,Chen-c}. Here the authors prove that Chua's system can be controlled by using a linear feedback controller and say that their numerical simulations show that simplest control can be achieved by perturbing only $V_{C_2}$. 

\subsection{Real data}
\subsubsection{Chua's circuit}
\begin{figure}[h]
\centerline{
\includegraphics[width=0.45\textwidth]{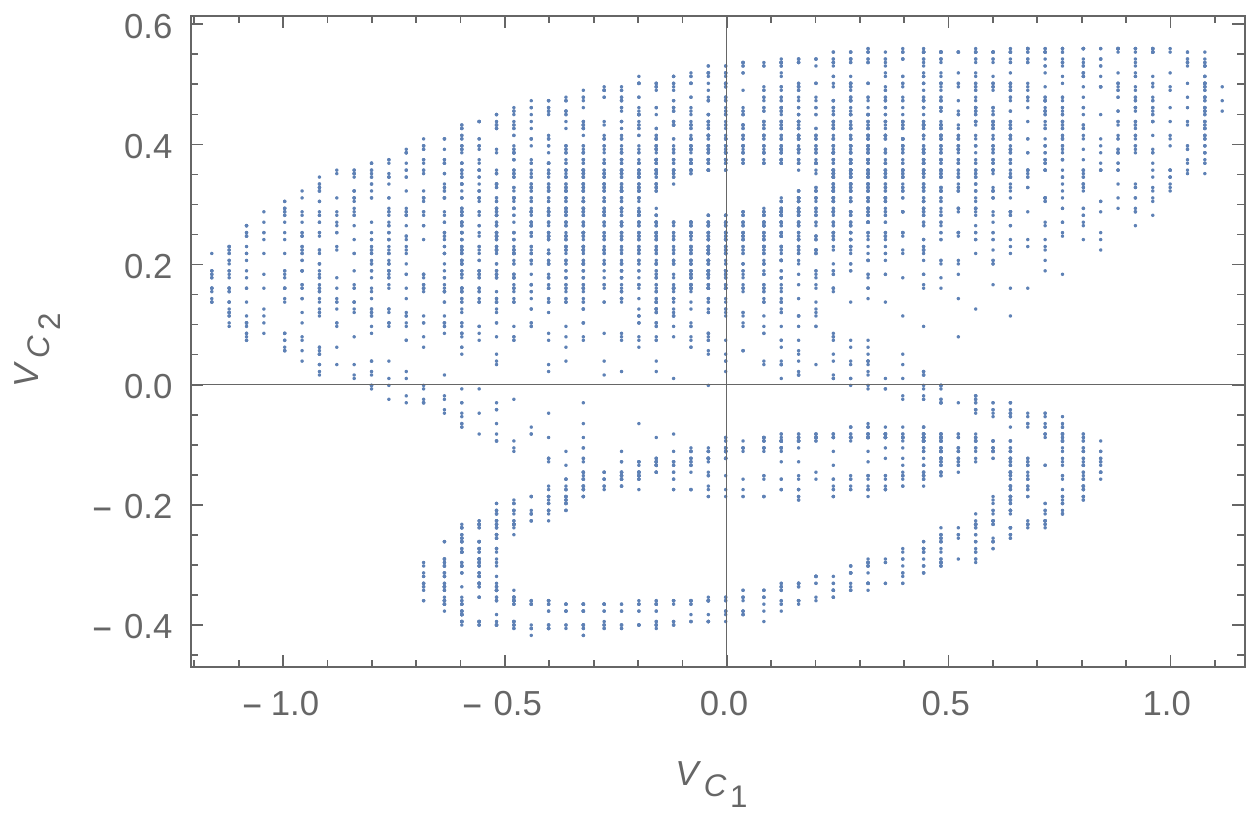}
\includegraphics[width=0.4\textwidth]{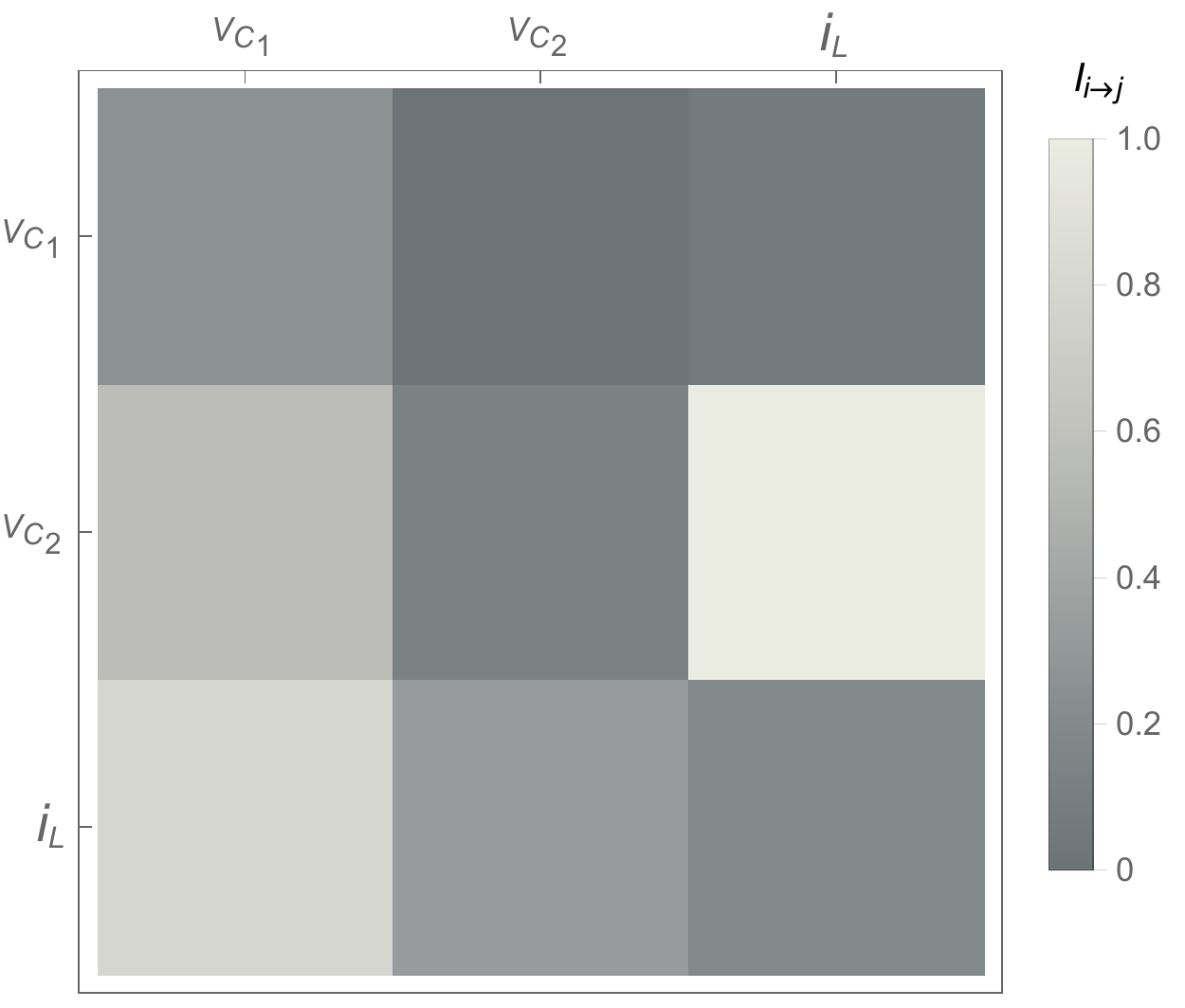}
}
\caption{Left: $V_{C_1}$ vs $V_{C_2}$, for 1024 experimental data points from Chua's circuit. Right: Information transfer between all pairs formed with signals $V_{C_1}$, $V_{C_2}$ and $i_L$. Here we chose the parameter $r$ as two times the standard deviation of the modeling errors at dimension three and the number of neighbors $k=1$, but the numerical experiments suggests that the results are not strongly dependent of this choice.}
\label{f4}
\end{figure}

In this case, an experimental implementation of the Chua's circuit was done, using operational amplifiers to simulate the Chua's  diode and the inductor $L$\cite{Gopakumar,Comment}. The current in the inductor was obtained by an indirect measure and the data, sampled with rate of $5 ms$, is shown in the left side of Figure \ref{f4}.

The right side of Figure \ref{f4} shows the information transfer estimated as explained before for all pairs formed with signals $V_{C_1}$, $V_{C_2}$ and $i_L$. 

In this case, $T_{1}=-0.268914$, $T_{2}=1.39525$ and $T_{3}=-1.09931$, which is consistent with the results obtained previously for the numerical simulation of this system.
 
\subsubsection{Physiological signals}
This experimental data correspond to time series taken from the Massachusetts General Hospital/Marquette Foundation (MGH/MF) Waveform Database in https://www.physionet.org/pn3/mghdb/ \cite{Welch}. From there, the series mgh003, mgh005, mgh006, mgh087, mgh097, mgh148, mgh160, mgh183, mgh190 and  mgh202, corresponding to electrocardiographic and blood pressure signals measured simultaneously, in patients in intensive care, were selected. To estimate the information transfer between both systems, $10000$ data points from each pair of series were used. In all cases  a greater transfer of information from arterial pressure or baro-reflex (br) system  to the electro-cardiac  (ec) system, $T=(I_{br \to ec} -I_{ec \to br})/(I_{br \to ec}+ I_{ec \to br})$, as obtained in \cite{Cisneros}.

\begin{figure}[h]
\centerline{
\includegraphics[width=0.65\textwidth]{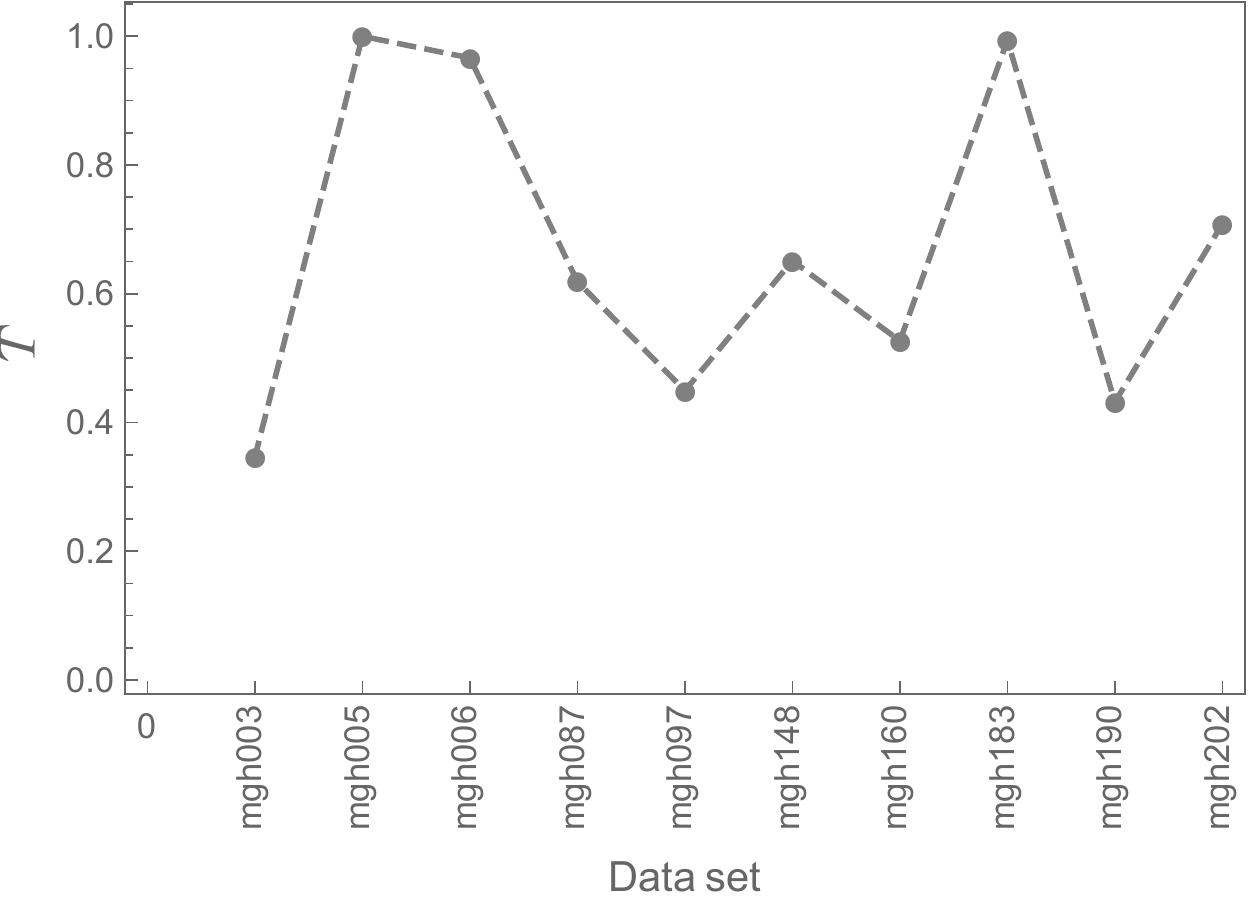}
}
\caption{Information transfer between pairs formed with electrocardiographic and blood pressure signals measured simultaneously. In this case $T$ represent the information transfer from the baro-reflex system to the cardiac system.}
\label{f5}
\end{figure}

\section{Final remarks}
Three aspects should be highlighted, regarding the performance of the proposed strategy, first, but not the 
most important, is referred to the low computational cost associated with the determination of the first nearest
neighbors compared, at least, to the cost to estimate CPD's using coarse graining or
calculating  correlation integrals as in \cite{Schreiber,Liebert}. Second, we must note that in the face of the results using experimental data, the estimation of the transfer of information based on nearest neighbors appears to be robust to the presence of moderate noise, given the consistency with the results in references \cite{Chen-a,Chen-b,Chen-c}. Although a more detailed study of these aspects it is needed to be conclusive, it is out of the scope of the present work and is subject of our current research. 

Finally, we emphasize that, although one of the advantages of Schreiber's estimator is that it is independent of 
models, and  with this proposal, it will become a dependent-model scheme, nevertheless we believe that the establishment
of a relationship between an useful probabilistic definition of information transfer and  deterministic modeling,
may be a practical idea in the development of methodologies to determine causal relationship between subsystems of complex 
systems or contribute in the design of control schemes for extended systems.

\bibliographystyle{plain}

\begin{thebibliography}{1}

\bibitem{Schultz} S. Schultz, F-C. Adochier, I-R. Edu, R. Schroeder, H. Costin, K-J.  Bar and A. Voss, Philos Trans A Math Phys Eng Sci,  {\bf 371}  (2013).

\bibitem{Liu}  Y-Y. Liu, J-J. Slotine, and A-L. Barabasi, Nature {\bf 473}, 167 (2011).

\bibitem{Wiener} N. Wiener, The theory of prediction. In Modern Mathematics for Engineer,  McGraw-Hill, 1956. 

\bibitem{Granger} C. Granger, Econometrica {\bf 37}, 424 (1969).

\bibitem{Schreiber} T. Schreiber, Phys. Rev. Lett. {\bf 85}, 461 (2000).

\bibitem{Bezruchko} B. Bezruchko, V. Ponomarenko, M. G. Rosenblum and A. Pikovsky,  CHAOS {\bf 13}, 179 (2003).

\bibitem{Ma} H. Ma, K. Aihara and L. Chen,  Scientific Reports  {\bf 4}, 1 (2014).

\bibitem{Ishiguro} K. Ishiguro, N. Otsu, M. Lungarella and Y. Kuniyoshi,  Phys. Rev. E. {\bf 77}, 026216 (2008).

\bibitem{Cisneros} L. Cisneros, J. Jim\'enez, M. G. Cosenza and A. Parravano, Phys. Rev. E.  {\bf 65}, 045204R (2002).

\bibitem{Rosenblum} M. G. Rosenblum and A. S. Pikovsky, Phys. Rev. E.  {\bf 64}, 045202R (2001).

\bibitem{Palus} M. Palus and A. Stefanovska, Phys. Rev. E.  {\bf 67}, 055201R (2003).

\bibitem{Vejmelka}  M. Vejmelka and M. Palus, Phys. Rev. E.  {\bf 77}, 026214 (2008).

\bibitem{Rubido} N. Rubido, A. Mart\'{\i}, E. Bianco-Mart\'{\i}nez, C. Grebogi, M. S. Baptista and C. Masoller, New Journal of Physics {\bf 16}, 093010  (2014).

\bibitem{Zhu} J. Zhu, J-J. Bellanger, H. Shu  and R. Le Bouquin Jeann\'es, Entropy {\bf 17}, 4173 (2015).

\bibitem{Vicente} R. Vicente and M. Wibral, In: {\it Efficient estimation of Information Transfer}, Springer-Verlag Berlin, 2014.

\bibitem{Ancona} N. Ancona, D. Marinazzo and S. Stramaglia, Phys. Rev. E. {\bf 70}, 056221 (2004).

\bibitem{Shannon} C. Shannon, The Bell System Technical Journal {\bf  27}, 379 (1948).

\bibitem{Palus-2001} M. Palus, Phys. Rev. E.  {\bf 67}, 046211 (2003).

\bibitem{Farmer} J. D. Farmer and J. J. Sidorowich, Phys. Rev. Lett. {\bf 58}, 845 (1987).

\bibitem{Garcia-1996} P. Garc\'{\i}a, J. Jim\'enez, A. Marcano and F. Moleiro,  Phys. Rev. Lett. {\bf 76},  1449 (1996).

\bibitem{Garcia-2007} P. Garc\'{\i}a and  A. Merlitti. The European Physical Journal Special Topics {\bf 143}, 261 (2007).

\bibitem{Loftgaarden} D. Loftgaarden and C. Quesenberry, Annals of Mathematical Statistics {\bf 36}, 1049 (1965).

\bibitem{Husmeier} D. Husmeier, In: {\it Perspectives in Neural Computing}, Springer-Verlag, 1999.

\bibitem{Takens}  F. Takens, In: {\it Dynamical Systems and Turbulence}, Springer Berlin, 1980.

\bibitem{Hasler} M. Hasler and Y-L. Maistrenko.  IEEE Trans. Circuits Syst. {\bf  44}, 856 (1997).

\bibitem{Chua} T. Matsumoto, IEEE Transactions on Circuits and Systems {\bf 31}, 1055 (1984).

\bibitem{Bollt} E. M. Bollt, International Journal of Bifurcation and Chaos {\bf 22}, 1250261 (2012). 

\bibitem{Chen-a} G. Chen and X. Dong, Journal of Circuits Systems and Computers {\bf  3}, 139 (1993). 

\bibitem{Chen-b} G. Chen, Chaos, Solitons \& Fractals {\bf 8}, 1461 (1997).

\bibitem{Chen-c} C. Chen and X. Dong,  Proc. of  IEEE  Inr. Symp. on Circ. Syst., Chicago, IL. 2604--2607 (1983).

\bibitem{Gopakumar} K. Gopakumar, B. Premlet and K. G. Gopchandran, International Journal of Electronics  {\bf 98}, 667 (2011).

\bibitem{Comment} The implementation of this circuit is already an undergraduate exercise, in this case it was done using an Educational Laboratory and Virtual Instrumentation Suite (NI-ELVIS) from National Instruments.

\bibitem{Liebert} W. Liebert, Phys. Lett. A. {\bf 142}, 107 (1989).

\bibitem{Welch} J. P. Welch, P. J.  Ford, R. S. Teplick and R. M. Rubsamen, The MGH/MF Waveform Database, 1992.

\bibitem{Cisneros} L. Cisneros and J. Jim\'enez,  Revista Mexicana de Física {\bf 49}, 17 (2003).

\end{thebibliography}

\end{document}